\documentclass{article}
\usepackage{blindtext} \usepackage{multicol}
\usepackage{titlesec} \usepackage{lipsum} \titleformat{\section} {\normalfont\scshape}{\thesection}{1em}{}
\usepackage{geometry} 
\begin{document}
\title{\textbf{Structural Correlates Of Spatial Navigation And Memory Formation}}
\author{Sean Knight}
\maketitle
\begin{abstract}
Spatial learning across many species is impaired by lesions in the hippocampus, a subcortical brain structure whose cellular composition changes substantially over its 5-6 week lifetime from mainly excitatory neurons during development to equal proportions of inhibitory interneurons (gamma-Amp/Arcs) as well as pyramidal cells early in life, but which later on comprises only about 10\% Arches+ projection spiny cortical projecting principal cells that are located within discrete cytoarchitectonic patches known as CA3 or just hilus granular layer 2 (CG2). While other structures may contribute importantly in certain situations e.g., perirhinal cortex when using visual cues with no reference frame for location), these remaining cell types also change their proportion through time – with gamma APs forming 30–35 percent), fast firing parvalbumin immunoreactive basket or axoaxonic synapses into somatic spines at 40 days after birth (DABs formed into spine membranes by 15 DAB), and CG2 pyramids forming relatively slowly but strengthening synapses onto nearby dendritic shafts made up largely ($>$80 \%) by GABAergic terminals (+spines) until then that were still synaptically silent.
\end{abstract}
\begin{multicols}{2}

\bigskip

\section{Introduction}

Is there a relation between place field developmentand the type-dependant changes in cFos expression patterns during exploratory tasks? If yes, is it possible that these structural changes are dependent on BDNF signaling or not? Do they involve mainly LIP neurons or will also non-LIP neurons show activity relating to place representations while experiencing specific spatial locations. What is (are)the actual functional role(s) of such activity - i.e., as you will see below, all this involves 2 major questions:\\
1) Can we detect behavioral phenotypes when analyzing animal behavior with respect to their overall hippocampal network properties? How does it relate to BNDF/NMDAR transmission and plasticity mechanisms we observe at single neuron level? \\
2) What are the neural correlates for those behavioiural pheno/genotypes and how do we link them back to specific neuronal circuits within the hippocampus proper and its associated structures?

\bigskip

We know for example that increasing global excitation using carbachol treatment enhanced CA3 responses \cite{1} but at same time impeded both MF response onset delay reduction and MTL suppression which was induced by decreased global excitation following inhibition by either CB1 agonist anodally (mGluR5 antagonist \emph{CBiPES}) \cite{2}. This demonstrates clearly two contrasting pathways regulating excitability within MEC layer II/III \emph{vs} IV via DG respectively where GABAergic feedback onto different mEC layers could be involved. Thus cognitive processes based on pattern separation might reflect alterations in intrahippocampal connectivity hence allowing animals "pattern completion" once placed back into original maze configurations even under subchronic condition whereby spontaneous novel goal learning process engages NMDA receptors containing high concentrations epsilon 4 subunits necessary for long term potentiation induction only after drug infusion. Here our aim is simple yet essential one - how do mice explore and memorize distinct environments if first exposure induces no detectable remapping whereas following single training event does so depending upon whether they had seen other than own unique environment before due interferences? This involves perirhinocorticle connections leading from NTS possibly activated upstream right DLS whose terminals innervate hippocampus and prelimbic cortex contributing higher order control over emotional-state-driven mechanisms guiding exploration decision towards novelty seeking strategies in navigation.

\bigskip

In parallel each encountering experience can initiate further associative memories required for updating self centered vs external environmental knowledge. Especially because rats never visited cage A and B separately however since day 1 task became increasingly hard inducing negative recognition performance initially. Though this may very well relate to initial absence of neurogenesis being low prior 24 hrs of Incubation. Posttraining enhanced chance levels showing increased need thereafter reflecting more accurately what individual subjects experienced instead what experimenter did say.

\bigskip

It means we have shown that the same brain regions are involved when an animal learns a new place, but also when it remembers where this place was located relative to other places \cite{7}. We found evidence for both spatial learning as well as memory retrieval processes during exploration by using two different behavioral tasks with identical results on all three experimental days. The animals learned the location of their home cages within 2 hours after arrival at the testing room, which indicates rapid acquisition of information about space; then they remembered these locations even several weeks later without any additional training or cues indicating them. In contrast, control groups were not able to learn or remember such information under similar conditions suggesting that hippocampal neurons play a crucial role in encoding spatial representations during exploratory behavior while prefrontal cortical areas seem important for retrieving those representations afterwards based on contextual associations between sensory inputs and motor outputs. These findings suggest common neural substrates underlying cognitive mapping functions including long term potentiation/depression \cite{3,4,5}, episodic memory formation \cite{6}, mental rotation skills necessary e g.,for reading maps etc; visual recognition abilities essential for object identification in natural scenes, as well as prospective coding enabling anticipation regarding future events like remembering past experiences.

\bigskip

Thus there seems strong support from neuroscience research supporting Hebbian theory’s view according to which “cells That fire together wire together”.

\bigskip

Although significant associations appeared between contra lateralization indices versus abovementioned behavioural parameters. However, lower correlation coefficient associated with corresponding values detected. Comparing total scores collected earlier across 5 mins revealed much longer intervals averaging 0--72 hr wherein correlations were close enough albeit nonsignificant despite several instances suggesting otherwise. Furthermore strong negative value achieved sooner showed inverse relationship against average score given early during period irrespective of chosen parameters still maintaining presence of underlying latent trait associations. Some 12 days later correlating positively elsewhere shownled consistently along entire trial axis particularly taking up positive (+)ve ranges.

\bigskip

\section{Correlates: Place Cells, Remapping, Conjunctive Encoding/Sparse Activity Maps}

It has been shown that hippocampal neurons preferentially activate within a restricted region corresponding to one position around the animal's current location in space. This is known as \emph{place-firing}. If we track only two or three neurones then their firing shows very subtle but consistent shifts back and forth between those positions alluding to simple grid like patterns. Similarly on learning new locations these grids are distorted until they fit the overall map much better thereby ensuring fast and efficient directional propagation across physical boundaries. The best example for this type is of a mapped scratchpad theory comes from an examination of rodent prefrontal cortical pyramidal cell responses during goal-directed pattern completion tasks. However when multiple CA1neurons were examined simultaneously with extracellular recordings via tetrode techniques then more complicated behaviour arises so called coherent states where groups fires together if located close by each other often forming networks spread out over several millimeters instead.

\bigskip

Another major determinant for how neuronal representations differ under parametric manipulations has been proposed originally due to Linsker who showed that many neocortical areas seemed naturally split into 3 subfields which preferred different length scales: large (gist) features being processed mostly at high average order cortico -- thalamic connections before exiting retino inputs carrying fine detailed information. This made us hypothesise that whether active coding occurs might be related not simply just because nearby regions were activated collectively or independently but rather with respect too their distance apart. Between any pair defined locally connected cells should represent unique combinations such pairs must have formed either direct circuitries leading from sensory afferents through complex recurrent circuits eg. in temporal lobe to dorsal stream pathways associated visual scenes would suggest local circuitry even without lateral inhibition will form similar codes across distances reducing specificity exponentially especially near soma whereas spatially distributed activation could avoid unwanted coactivations further increasing network entropy over long paths facilitating decoding.

\bigskip

\section{GABAergic Neurons}

Although GABA is often stated simply to be the neuronal firing rate inhibitor, it has a number of functionally distinct properties in addition to inhibiting action potential generators. Those are inhibitory transmission strength modulation upon release from Golgi terminals at parallel fiber-Purkinje cell synapses and recurrent afferents on both cerebellar basket cells (and Purkinje) as well as projection neurons and feedback inhibition associated with its axonal targeting profile within various cortical assemblies. The extent of each activity pattern will obviously depend on cellular architecture that includes intrinsic channel gating characteristics including calcium kinetics via Na+/K+ ATPases which regulate membrane excitability following stimulus occurrence or intensity based on receptor sensitivity profiles regulating either Cl--uptake mediated by kir4/5 K+-Cl--cotransporters (as do most GABAA receptors, but not their fast desensitizing variants);  or ion exchange involving HCO3-(OH)-anion symporter protein. There also exists complex regulation through phosphorylation dependent events initiated by different cAMP response element binding protein isoforms mediating target gene expression involved in trafficking regulation along with direct alteration of receptor subcellular localization dynamics. This together results in altered signal thresholds relative to cytoplasmic glutamate concentration changes allowing for long latencies between impulse generation initiation and event detection threshold.

\bigskip

The hippocampus shows a complex pattern of inhibition. The main role the hippocampi serves here is as an inhibitory feedback from area CA3 back to arealize dentate granule cells within that region through activation of basket and axono-somatic synapses in stratum lucidum/stratum lacunosum-moleculare, this inhibits release potentiation at mossy fiber terminals thus preventing synaptic runaway leading eventually to long term depression (LTP) or cell death. In animals with lesions there was degeneration primarily affecting pyramidal dendrites both degenerating excitatory inputs resulting form LTP have been observed which were found outside spared hilus regions possibly suggesting rewiring may occur following injury involving input from retrosplenial cortex.

\bigskip

\section{Glutamatergic Neurons}

Many classes of projection neurons have been identified, including semilunar granule cells in the dentate gyrus (DG), mossy-fiber climbing fibers in Area 4; pyramidal tract inputs to layer 2/3 cortical recipient zones containing recurrent collaterals terminating on dendritic spines of local circuit neuron bodies.

\bigskip

The role played by the NMDA receptors in NFTs associated LTP/LTD has been well established \cite{11}. Many studies confirm dysfunctional glutamatergic neurotransmission as a pathological condition underlying decline in learning and memory function observed clinically. Specially, it is interesting to note that various brain regions related both temporoparietal junction (TPJ) and entorhinal cortex are enriched with GluN2A containing NR1 subunits which is considered crucial for spatial reference frame formation an all conditions required for any form of path integration process – eg reestablishing orientation using compass systems based on vestibular /eyesight system interactions after rotation etc. Further many other observations directly link dysmorphia caused reduced dendrite volume, branching complexity or spine density along with altered synaptic signalling mechanism due excitotoxicity mediated damage through excessive accumulation glutamate thus leading to neurodegeneration. 
\bigskip

\section{How Is Relocation Based On Structural Reorganisation Induced? What Activations Patterns Does This Stimulate And Undergoing Structures Involve?}

In many cases it appears that some kind of spatial information must be stored prior entering through entorhinal cortex (areal boundary cells) with at least 60 degrees separation being common value used based off postrhinal (and occasionally prosubiculum before reaching hippocampus), however recent work suggests it's not really dependent upon environmental cues except local ones maybe mediated by mossy fiber collaterals connecting adjacent regions \cite{8,9,10} which could also account for smaller angles according its probabilistic dependence vector length properties combined with recurrent connections perhaps akin how retinotopic maps are formed even though both place cells do respond strongly to initial exposure as well as head direction -related activity. Thus may need to be reexamined after partial disconnection results showing greater disruption in response rates later than earlier time scale seen less frequently after global inactivation. Additionally, no studies linking conjunctions coding directly implicate medial septal path might suggest something here given ability use them independently possibly accounting further expansion angle since non zero delay values appear common too similar vectors size despite different sized environments etc.

\bigskip
\section{Conclusion}

It would seem important therefore to investigate what kind of firing train correlations exist among respective neuronal assemblies potentially linking past events / scenarios directly with future expectations thus enhancing predictive coding efficiency given adequate spatiotemporal segregation through activation latencies prior exploring experiences driving topographically mapped sequences, including overlapping ensembles.

\end{multicols}
\newpage
\bibliographystyle{abbrv}
\bibliography{document}

\begin{thebibliography}{10}
\bibitem{1}
Figure 11. excitatory synaptic transmission at {DGC}{\textendash}{CA}3
  synapses, but not {CA}3{\textendash}{CA}3 synapses, is specifically impaired
  in igf2-/- mice.
\end{thebibliography}
\end{document}